\documentclass[12pt]{article}

\usepackage[english]{babel}

\begin{document}

\begin{center}
{\bfseries SPIN AS AN ADDITIONAL TOOL FOR QGP INVESTIGATIONS}

\vskip 5mm

J.~Karachuk, S.S.~Shimanskiy$^{\dag}$

\vskip 5mm

{\it JINR, Dubna }

\vskip 5mm

( $\dag$ {\it E-mail: shimansk@sunhe.jinr.ru })
\end{center}

\vskip 5mm

\begin{center}
\begin{minipage}{150mm}
\centerline{\bf Abstract} \vskip 5mm

\ \ \ \ \ The nearest two years on experiment STAR the upgrade is
planned, which will make it possible to identify particles up to
momentum $\sim$ 3 GeV/c. This will open possibility to carry out
new and more detailed researches of properties of a nuclear matter
formed in nucleus-nucleus collisions at RHIC. In this work we
offer to carry out of the polarization studies, which can give
important additional information about the process of forming the
new state of nuclear matter, and also about properties of the
formed state.

\ \ \ \ \ A unique probe of information about all stages of
formation and evolution of nuclear matter are dileptons, due to
their electromagnetic interaction with the nuclear matter. In this
work we pay main attention to the examination of polarization
characteristics of dileptons.
\end{minipage}
\end{center}

\vskip 10mm

\section{Introduction}

The first studies on the accelerator RHIC (BNL) have shown that
properties of the nuclear matter formed in Au-Au-collisions
substantially differ from those of the nuclear matter formed at
collisions of nuclei on the accelerator SPS (CERN) \cite {bib1}.
It made it necessary to reconsider theoretical views on the
properties of formed quark-gluon plasma \cite {bib2}. Up to the
present time, the search of unequivocal signatures of formation of
quark-gluon plasma remains active. From our point of view,
polarization studies give an additional tool allowing to detect
the formation of new states of nuclear matter.

The upgrade of STAR setup \cite {bib3} will allow one to identify
particles with momenta up to $ \sim 3~GeV/c $. In particular, it
will allow us to investigate dilepton production in the range of
effective masses \ \ $m _ {e ^ + e ^-} <5~GeV $. It is exactly in
this range of effective masses that the majority of particles
produced by quark-gluon plasma are expected. Dileptons and photons
are unique probes because they interact with  nuclear matter only
via electromagnetic interaction. Due to their weak interaction
dileptons and photons carry information about all stages of
nucleus-nucleus collisions without noticeable distortion. This is
a principal  distinction of photons and leptons from hadron
probes. Comparison of characteristics received by electromagnetic
and hadrons probes will allow us to investigate properties of the
forming nuclear matter.

It is expected that in nucleus-nucleus collisions the quark-gluon
plasma(QGP) is formed, which is essentially a new source of
secondary particles. Characteristic features of this new source,
which we attempt to find as signals of QGP formation, are the
object of studies. The thermalization may be the main feature of
the QGP.

The thermalization means that the information about initial states
of nucleus-nucleus collisions has been lost (for example, the
information about the initial direction of collisions is lost due
to multiple secondary interactions). As a consequence we should
see disappearance of all types of polarization connected with the
direction of initial state for particles produced by the plasma
source. The polarization for particles produced by the plasma can
depend only on hadronization characteristics and the plasma
collective motion. Therefore, there should be no transverse
polarization or longitudinal polarization connected with the
direction of initial collision. That is why we can use
polarization studies as an additional important information to
identify the QGP formation.

But it is very important to determine the processes and energy
ranges where the absence of polarization can be regarded as a
proof of the QGP formation. For the first time, this type of
possibility has been proposed and discussed in detail about ten
years ago. In 1994 one of the authors (S.S.S.) proposed to
CERES/NA45 collaboration to carry out polarization studies of low
mass dileptons ($0.2~<~m_{ee}~<0.6~GeV/c^2$) to explain the nature
of dilepton enhancement in the nucleus-nucleus collisions.
However, further analysis showed that CERES/NA45 had a very narrow
acceptance \cite{bib4}. Nowadays there are new setups with
possibilities to carry out polarization investigations of
dileptons to find formation of the QGP phase in nuclear-nuclear
collisions. That is why we think that it is important to analyze
the possibility for polarization studies now. Some experiments
have been carried out for some years (HADES, NA60, STAR and
PHENIX).


\section{Polarization and thermalization}

CERES/NA45 studies \cite{bib5} of the dilepton production in
nucleus-nucleus collisions with nuclear beams at SPS(CERN) have
shown an enhancement of the dilepton production in the mass region
$0.2~GeV < m_{e^+e^-} < 0.8~GeV$. If this enhancement come from
the thermalalized source, we should not see any dilepton
anisotropies. That is not so for a secondary $\pi^{+} \pi^{-}$ -
annihilation process where a strong anisotropy of the electron
(positron) emission should occur. Moreover, there must be energy
dependence of this anisotropy which is opposite to a thermalized
source case which has no the energy dependence. Therefore, we have
a real possibility to distinguish  these two subprocesses. That
time there were no quantitative theoretical estimations of these
effects. The author (S.S.S.) had asked theoreticians from JINR for
theoretical examination these effects \cite{bib6}. This work was
continued in collaboration  with a theoretical group of Giessen
\cite {bib7, bib8}. These are the only quantitative theoretical
predictions up to now. Theoretical predictions for the anisotropy
of leptons in the region of small masses and for energies from
SIS(GSI) to SPS(CERN) were based on the Hadron-String
Dynamics(HSD) model \cite{bib9}. At that time there was no
possibility to carry out studies in the region of the dilepton
masses $m_{e^+e^-} > 1~GeV$ that is why all predictions had been
limited to the region of small masses.

Theoretical studies have shown that the dilepton polarization
characteristics allow one to separate different subprocesses (and
models) not only in nucleus-nucleus collisions but in
nucleon-nucleon interactions. In 1998 CERES/NA45 has obtained the
data for $p_T$ dependences  of the dilepton pair production
\cite{bib10}. These data have shown that the dilepton enhancement
comes from a low-$p_T$ region of pairs. Theoretical description of
this enhancement for $p_T\sim 0~GeV$ \cite{bib11} remain difficult
and can be partly explained by a combination of a strong
polarization with the narrow CERES/NA45 acceptance. It implies the
domination of the process by the annihilation of nonthermalized
pions. May be this is the first indirect proof which tells us that
the main source that is crucial for the dilepton enhancement is
not a thermalized source. If it is so, we have the first direct
observation of the annihilation process of secondary pions and its
studies have some independent interest.

RHIC has opened essentially new opportunities for polarization
studies of dileptons. It is the possibility to carry out
investigation of the thermalized source in a wider range of
dilepton masses and immediately in the center-of-mass system. The
article \cite{bib12}  examines the contribution of the thermalized
dilepton source for energies SPS and RHIC. At these energies there
is additional interesting region of masses $1~GeV < m_{e^+e^-}
<5~GeV$,  where the thermalized source of dileptons competes with
the Drell-Yan annihilation. As well as in a case of small dilepton
masses, we have competition of two subprocesses: the thermalized
source and the Drell-Yan process. The latter gives a strong
alignment of virtual photons and as a consequence, of the
anisotropy in the lepton angle distribution. Different energies of
nucleus-nucleus collisions, the possibility to select  events with
different impact parameters make a real possibility to
unambiguously reveal the appearance of the thermalized source. In
our opinion such studies now are unique in the sense that they
allow us to detect directly the occurrence of a thermalized source
of particle production.

There are independent reasons to study the polarization
characteristics of dileptons in nucleon-nucleon collisions at RHIC
for small and middle mass regions with polarized proton beams.
These investigations will give an additional information about
polarized structure of nucleons in the region $x\sim 10^{-2}$.

RHIC setups have possibility to detect hadrons, that is why we can
propose to compare hadron and dilepton polarization
characteristics in nucleus-nucleus collisions to investigate
influence of formed nuclear matter. For example, the comparison
polarization characteristics of dileptons and $\pi^{+}
\pi^{-}(K^{+}K^{-})$ - pairs gives new information about the QGP
formation. There are many other hadron probes where polarization
should be used to prove the QGP formation. We can propose to
investigate the polarization of $\Lambda$ which comes from
$\Lambda K^{+} (\overline{\Lambda} K^{-})$- back to back pairs.

Till now we discuss the so-called regions of continuous dilepton
mass spectrum. The resonance regions are very  interesting too.
But we do not have so many quantitative predictions. The
$\rho$-meson alignment can serve as a sensitive tool to study
mechanisms of formation \cite{bib13}. Article \cite{bib14}
proposed to investigate $J/\psi$ polarization in nucleus-nucleus
collisions. The QGP formation should give a huge value of the
alignment for $J/\psi$ in contrast to nucleon-nucleon collisions
where the alignment must be zero. It is very useful to compare
characteristics of resonance polarizations from dileptons(photons)
and hadrons decay modes.

\section{Conclusion}

In this report we have discussed some ideas which allow one to
receive the new additional experimental information which can help
to prove the QGP formation. Polarization  \\
characteristics are a thin tool, and these experimental data will
give possibility to speak about the detection of the thermalized
state unequivocally.  The authors are grateful to E.L.~Bratkovskay
and O.V.~Teryaev for interest and support.



\begin{thebibliography}{99}

\bibitem{bib1}
U.~Heinz, M.~Jacob, Evidence for a New State of Matter: An
Assessment of the Results from the CERN Lead Beam Programme,
{\bfseries nucl-th/0002042},  (2000). \\
J.~Adams et al, Experimental and Theoretical Challenges in the
Search for the Quark Gluon Plasma: The STAR Collaboration’s
Critical Assessment of the Evidence from RHIC Collisions,
{\bfseries nucl-ex/0501009},  (2005).

\bibitem{bib2}
E.~Shuryak, A strong coupled quark-gluon plasma, J.Phys. G:
{\bfseries 30}, S1221 (1901). \\
M.~Gyulassy, L.~McLerran, New form of QCD matter discovered at
RHIC, Nucl. Phys.{\bfseries A 750}, 30 (2005).

\bibitem{bib3}
T.~Hallman, The STAR Decadal Plan, RHIC Open Planning Meeting,
December 3-4, 2003.

\bibitem{bib4}
Guy\ Tel-Zur,\ Ph.D.\ thesis,\ pp.70,71,73 (1996). \\
http://www.physi.uni-heidelberg.de/physi/ceres/theses.html

\bibitem{bib5}
G.~Agakichiev et al., Enhanced Production of Low-Mass Electron
Pairs in 200~GeV/Nucleon S-Au at the CERN Super Proton
Synchrotron, Phys.Rev.Lett.,{\bfseries 75}, 1272 (1995).

\bibitem{bib6}
E.L.~Bratkovskaya, O.V.~Teryaev, V.D.~Toneev, Anisotropy of
dilepton emission from nuclear collisions, Phys.Lett. {\bfseries
B~348}, 283 (1995).

\bibitem{bib7}
E.L.~Bratkovskaya et al., Anisotropy of dilepton emission from
nucleon-nucleon interactions, Phys.Lett. {\bfseries B~348}, 325
(1995).

\bibitem{bib8}
E.L.~Bratkovskaya, W.~Cassing, U.~Mosel, Probing hadronic
polarizations with dilepton anisotropies, Z.Phys. {\bfseries
C~75}, 119 (1997).

\bibitem{bib9}The Hadron-String Dynamics (HSD) Model, \\
{\bfseries
http://www.th.physik.uni-frankfurt.de/$\sim$brat/hsd.html}

\bibitem{bib10}
G.~Agakichiev et al., Low-mass $e^+e-$ pair production in 158 A
GeV Pb-Au collisions at the CERN SPS, its dependence on
multiplicity and transverse momentum, Phys.Lett.{\bfseries B~422},
405 (1998).

\bibitem{bib11}
W.~Cassing, E.L.~Bratkovskaya, Hadronic and electromagnetic probes
of hot and dense nuclear matter, Phys.Report {\bfseries 308}, 65
(1999).

\bibitem{bib12}
R.~Rapp, Dilepton in high-energy heavy-ion collisions, PRAMANA,
{\bfseries vol.60,\ No.4}, 675 (2003)

\bibitem{bib13}
A.V.~Efremov, O.V.~Teryaev, On Vector Meson Spin Alignment in
Fusion Model, {\bfseries P2-82-832}, Preprint of JINR, Dubna 1982.

\bibitem{bib14}
B.L.~Ioffe, Charmonium polarization in $e^{+}e^{-}$ and heavy ion
collisions, X Advanced Research Workshop on High Energy Spin
Physics (NATO ARW DUBNA-SPIN-03), p.79, Dubna 2004. {(\bfseries
hep-ph/0310343)}

\end{thebibliography}
\end{document}